\documentclass[final,3p,times,twocolumn]{elsarticle}

\usepackage{subcaption}
\usepackage{hyperref}
\usepackage{amsfonts}
\usepackage{amssymb}
\usepackage{amsmath}

\usepackage{arydshln}
\usepackage{sidecap} 

\journal{Nuclear Inst. and Methods in Physics Research, A}

\begin{document}

\begin{frontmatter}



\title{Coherent radiation in axially oriented industrial-grade tungsten crystals: a viable path for an innovative $\gamma$-rays and positron sources}


\author[infnferrara]{N. Canale \corref{first}} 

\author[ferrara]{M. Romagnoni\textsuperscript{\dag ,} \corref{coauthor1}} 
\author[infnferrara]{A. Sytov\textsuperscript{\dag, } \corref{coauthor2}} 

\author[IJCLAB]{F. Alharthi}
\author[frascati]{S. Bertelli} 
\author[como,bicocca]{S. Carsi}
\author[IJCLAB]{I. Chaikovska}
\author[IJCLAB]{R. Chehab}
\author[padova,legnaro]{D. De Salvador}
\author[infnferrara,ferrara]{P. Fedeli} 
\author[infnferrara,ferrara]{V. Guidi} 
\author[]{V. Haurylavets} 
\author[como,bicocca]{G. Lezzani}
\author[infnferrara]{L. Malagutti} 
\author[como,bicocca]{S. Mangiacavalli}
\author[ferrara]{A. Mazzolari}
\author[trieste]{P. Monti-Guarnieri}
\author[infnferrara,ferrara]{R. Negrello} 
\author[infnferrara]{G. Paternò \corref{corr}} 
\author[gransasso]{L. Perna}
\author[como,bicocca]{M. Prest} 
\author[como,bicocca]{G. Saibene}
\author[como,bicocca]{A. Selmi}
\author[legnaro]{F. Sgarbossa}
\author[frascati]{M. Soldani} 
\author[]{V.V. Tikhomirov}
\author[padova,legnaro]{D. Valzani}
\author[bicocca]{E. Vallazza}
\author[como,bicocca]{G. Zuccalà}

\author[infnferrara]{and L. Bandiera \textsuperscript{\ddag ,} \corref{last}}

\cortext[first]{First author: \textit{Email address: ncanale@fe.infn.it}}
\cortext[corr]{Corresponding author: \textit{Email address: paterno@fe.infn.it}}
\cortext[coauthor1]{\textsuperscript{\dag}Co-author: \textit{Email address: marco.romagnoni@unife.it}}
\cortext[coauthor2]{\textsuperscript{\dag}Co-author: \textit{Email address: sytov@fe.infn.it}}
\cortext[last]{\textsuperscript{\ddag}Last author: \textit{Email address: bandiera@fe.infn.it}}

\affiliation[infnferrara]{organization={INFN, Sezione di Ferrara},  city={Ferrara}, state={Italy}}
\affiliation[ferrara]{organization={Università degli Studi di Ferrara},  city={Ferrara}, state={Italy}}
\affiliation[IJCLAB]{organization={Université Paris-Saclay, CNRS/IN2P3, IJCLab},  city = {Orsay}, state = {France}}
\affiliation[frascati]{organization={INFN, Laboratori Nazionali di Frascati},  city={Frascati}, state={Italy}}
\affiliation[como]{organization={Università degli Studi dell'Insubria},  city={Como}, state={Italy}}
\affiliation[bicocca]{organization={INFN, Sezione di Milano Bicocca},  city={Milano}, state={Italy}}
\affiliation[padova]{organization={Università degli Studi di Padova},  city={Padova}, state={Italy}}
\affiliation[legnaro]{organization={INFN, Laboratori Nazionali di Legnaro},  city={Legnaro}, state={Italy}}
\affiliation[trieste]{organization={Università degli Studi di Trieste, INFN, Sezione di Trieste},  city={Trieste}, state={Italy}}
\affiliation[gransasso]{organization={Gran Sasso Science Institute, INFN, Laboratori Nazionali del Gran Sasso}, city={L'Aquila}, state = {Italy}}

\begin{abstract}
\noindent It is well known that the alignment of an high-energy electron beam with specific crystal directions leads to a significant increase of the coherent radiation emission. This enhancement can be exploited to create an intense photon source. An elective application is an innovative positron source design for future lepton colliders. Such scheme takes advantage of lattice coherent effects by employing a high-Z crystalline radiator, followed by an amorphous metallic converter, to generate positrons via a two-step electromagnetic process. Additional applications can be in neutron production through photo-transmutation and radionuclide generation via photo-nuclear reactions.
In this work, we present experimental results obtained from beam tests at CERN's PS facility using commercial industrial-grade tungsten crystals. The obtained results demonstrate the robust performance of industrial-grade radiators, even with their inherent imperfections, suggesting that it is possible to simplify the supply process and it is not strictly necessary to rely on highly specialized research infrastructures.
\end{abstract}



\end{frontmatter}



\section{Introduction}
\subsection{Coherent effects in oriented crystals} \label{sect:coherent_effects}
\noindent When charged particles move inside a crystal at small angles relative to atomic rows or planes, coherent effects manifest \cite{Kumakhov76}. Namely, the particles experience correlated interactions with neighbouring lattice atoms. Therefore, their dynamics can be described by considering the continuous potential of atomic strings/planes. For angles smaller than the Lindhard critical angle, $\theta_L = \sqrt{2U_0/E}$, where $U_0$ and $E$ are respectively the potential of the considered planes or axis and the energy of the beam, the particles are in channeling condition \cite{lindhard1965influence}. Namely, they are transversely trapped in a potential well that forces them to move far or close to strings of atoms of the crystal, depending on their charge. Due to the resulting quasi-periodic motion, the particles emit intense electromagnetic radiation, exceeding that from randomly oriented crystals \cite{Baier-Katkov1986}. A further radiation mechanism, namely the coherent bremsstrahlung, plays an important role far from channeling condition. In this case, the particles that impinge at a small angle $\theta > \theta_L$ with respect to a lattice symmetry (plane or axis) periodically cross strings of crystal atoms, feeling their intense potential. This gives rise to constructive interference of the bremsstrahlung emission when the momentum transferred to the crystal nuclei matches a lattice reciprocal vector \cite{saenz1991theory}. The resulting radiation is dipole-like with an energy spectrum characterized by hard and intense peaks, provided that the multiple scattering effects are not strong enough to completely smear them.  When negatively charged particles, as electrons, impinge at small angles with respect to an axis of a thick high-Z crystal, they are mostly found in an over-barrier state and move chaotically along the axis. This is due to the intensity of the axial potential, and the increased probability of scattering with the crystal high-Z nuclei. This condition holds true also for imping angles $\theta < \theta_L$\footnote{For a tungsten crystal aligned along its $\langle111\rangle$ axis and a 6 GeV electron beam $\theta_L \approx 0.54$ mrad.}, since the particles initially channeled soon dechannel, due to the aforementioned scattering contribution. The radiation emission assumes intermediate characteristics between the pure channeling radiation and the coherent bremsstrahlung, with a continuous energy spectrum characterized by a significant enhancement compared to standard Bethe–Heitler bremsstrahlung \cite{EPJC22}. For $\theta >> \theta_L$, the coherent bremsstrahlung dominates. 

\subsection{Crystal-based positron sources: theoretical concepts}

\noindent We can take advantage of the radiation due to coherent effects in a crystal to realize intense positron sources based on the process of photon conversion into electron-positron pairs. Indeed, a crystal-based positron source was first proposed in 1989 \cite{Chehab89}. Axial orientation in high-Z metallic crystals is preferred over planar one since the radiation emission is more intense, due to the higher potential values, and since it is mainly composed of soft photons at the energy range of interest for the primary electrons (few GeV) \cite{Artru94}. Indeed, these photons convert in the target mostly into positrons with an energy lower than 100 MeV, which are typically more efficiently accepted within the capture section of an injector \cite{ALHARTHI25}. Also, the angular range within coherent effects manifest in the case of axial alignment is wider, up to several mrad.
Future lepton colliders are exploring the "hybrid scheme" \cite{EPJC22,Chehab2002,Chaikovska22,Soldani24}, which takes advantage of a thin crystal radiator (thickness $<$ one radiation length $X_0$ \cite{PDG}) to generate photons that convert into $e^+e^-$ pairs in a downstream converter amorphously oriented (Fig.~\ref{fig:1}.c).
This setup offers key advantages over single stage sources: the energy deposition is distributed between two objects. Since the radiator thickness is typically smaller than $X_0$, the shower development inside it is negligible and, therefore, the total energy deposited and the peak energy deposition density (PEDD) are small. Also, a thinner target than for the conventional scheme (Fig. \ref{fig:1}.a) is required to obtain the desired positron production yield \cite{EPJC22}. This makes the total energy deposited inside the target, and hence the thermal load, smaller. Furthermore, due to the radiator-converter distance, the fingerprint of the beam impinging on the converter increases with respect to the single stage case, with a consequent flux density reduction. This reduces the PEDD in the target without a corresponding decrease of the e+ yield, which, on the contrary, would occur by simply reducing the drive electron beam current \cite{Soldani24}. A single thicker oriented crystal (Fig. \ref{fig:1}.b) is a simpler and more compact configuration, however, in this case, the PEDD may be very high and the heat loads cause more severe problems (thermic stress) for a crystal target than for an amorphous one, due to more complicated thermic contacts required in working condition. For this reason, the latter approach may be advantageous only for not so intense sources, such as the one needed by FCC-ee \cite{Chaikovska22}. On the other hand, for the extremely intense sources typically required by linear colliders, advanced hybrid schemes involving collimators or bending magnets can be taken into consideration. Indeed, collimators can be used to cut the tails of the photon beam emitted by the radiator crystal, thus reducing the deposited power in the converter. Besides, we can take advantage of magnetic fields to remove the charged particles from the beam impinging on the converter and reduce the PEDD \cite{Soldani24}.

\subsection{Crystal-based positron sources: experimental and simulation studies}

\begin{figure}[h]
    \centering
    \includegraphics[width=0.8\linewidth]{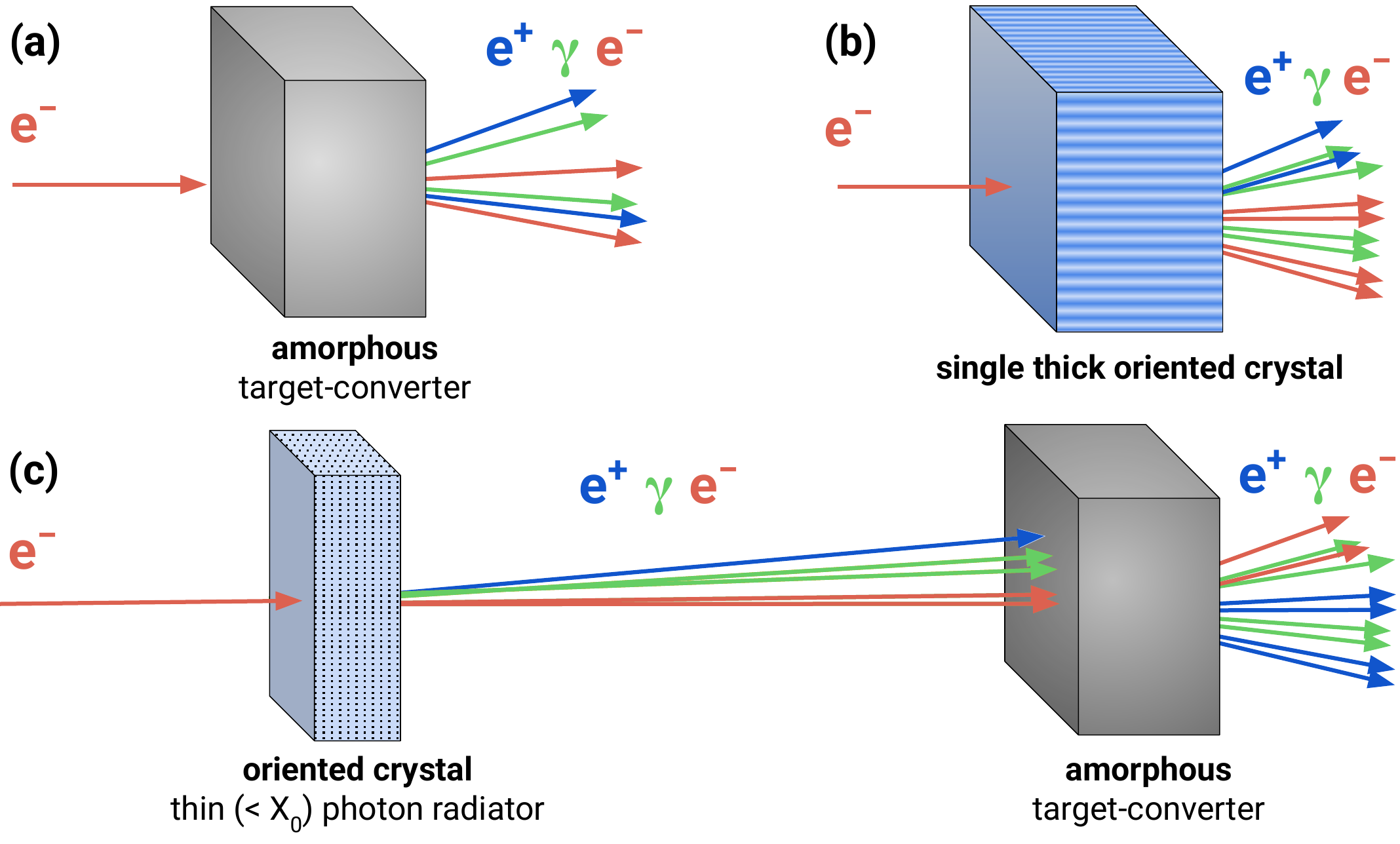}
    \caption{Comparison of various hybrid positron source schemes: (a) conventional source, (b) single thick crystal, (c) basic hybrid design.}
    \label{fig:1}
\end{figure}

\noindent Several experiments at CERN, DESY, KEK, MAMI, and Orsay investigated crystal-based positron sources considering various crystalline materials and thicknesses \cite{EPJC22,Chehab2002,Soldani24,Artru96,Suwada03,Bent_W}. Proof-of-principle tests at Orsay with a 2 GeV electron beam confirmed the feasibility \cite{Artru96}, showing that thin (1–2 mm) crystals produce high photon rates while limiting the heat load. 
As previously said, in a hybrid scheme an electromagnetic shower develops in the converter as a consequence of the impinging photon beam. The shower peaks at the downstream exit and causes high gradients in the energy deposition distribution.
Studies at LANL and LLNL indicated that a PEDD exceeding 35 J/g may cause target failure \cite{Maloy01, Stein01}, establishing this value as the safety threshold for tungsten targets in positron sources.
While, the \textit{WA 103} experiment showed that thick crystal (Fig. \ref{fig:1}.b) positron yields can meet linear collider requirements \cite{Suwada03}, heat load issues persist. 
However, as a supplement to what was previously said, recent simulation studies using Baier-Katkov's method \cite{Baier-Katkov1986} suggest that, for moderate-intensity beams, a single thick crystal can be effectively and safely used \cite{ALHARTHI25}. This approach offers advantages over conventional methods while maintaining safe PEDD levels.

\subsection{Industrial-grade radiators}

\noindent In this paper, we present an experimental investigation of the radiation due to coherent effects in oriented crystals generated by a 6 GeV electron beam passing through different high-density oriented crystalline samples (\textit{i.e.}, 1.5 and 2 mm tungsten crystals aligned with respect to $\langle111\rangle$ axis). Specifically, we measured the total energy loss due to the radiation emitted by electrons crossing the crystals. A portion of this radiation was converted into electron-positron pairs by a thin copper layer in order to be directly detected and counted.

Similar measurements were performed at DESY \cite{EPJC22} using $\langle100\rangle$ tungsten crystal manufactured by a research-centre. While this crystal exhibited high crystallographic quality, it had a sub-optimal orientation. Additionally, the beam energy used in \cite{EPJC22} was 5.6 GeV, whereas the experiments presented here employed a 6 GeV electron beam, consistent with the baseline energy for the FCC-ee injector CDR \cite{ChaikovskaIPAC19,CDR-FCC}. 

The results presented in this paper represent the first measurements conducted with commercially available tungsten crystals, emphasizing their potential for practical applications.

\section{Test Beam at CERN PS for crystal characterisation}
\noindent The study was carried out at the T09 beamline of the CERN Proton Synchrotron beam test facility \cite{CERNPS}, using mainly a 6 GeV electron beam.

\subsection{The crystalline samples}
\noindent Two tungsten crystals, both aligned along the $\langle$111$\rangle$ axis, with thicknesses of 1.5 mm and 2 mm (0.43 and 0.57 $X_0$) and transverse area of $\sim$ 1 cm$^{2}$, respectively, were tested. This crystallographic axis is known to exhibit the strongest potential for a body-centred cubic (BCC) lattice, which characterizes tungsten.

\begin{figure}[ht] 
    \centering 
    \includegraphics[width=0.95\linewidth]{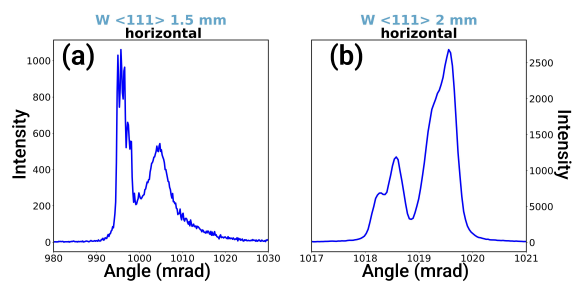}
    \caption{XRD rocking curves for tungsten crystals of 1.5 mm thickness (a) and 2 mm thickness (b). Broader peaks reflect strain, mosaicity, or lattice imperfections, while multiple peaks highlight the presence of distinct crystalline domains—features characteristic of industrial-grade crystals.} 
    \label{fig:2}
\end{figure}

The lattice quality of the samples was evaluated using X-ray diffraction, specifically assessing their mosaicity (\textit{i.e.} the average angular spread of crystallites within the sample). This is a critical parameter as it significantly influences performance due to the strong angular dependence of the lattice coherent effects. The measurements were carried out using a High-Resolution X-Ray Diffraction (HR-XRD) at the Ferrara University laboratories, employing photons with an energy of 8.04 keV ($K\alpha_1$ line of Copper).
The analysis of the X-rays diffraction intensity as the crystal was rotated around the Bragg angle allowed to determine the mosaicity at each position on the sample surface. The results of this characterization for the Tungsten crystals are showing in Fig. \ref{fig:2}.
In the rocking curves presented here (\textit{i.e.} the intensity of X-ray diffraction as a function of the angular deviation of the crystal relative to the incident beam), the peak widths range from 1 to 10 mrad, with multiple peaks indicating the presence of distinct crystalline domains.
The HR-XRD probe only the surface structure, revealing significant mosaicity, while the quality of the crystal bulk remains unclear. However, if multiple crystal domains are present, alignment to the domain with the strongest coherent effects is possible, as demonstrated for the 2 mm tungsten crystal, where the dominant domain was successfully aligned to optimize performance.
In contrast, the research-grade crystal described in \cite{EPJC22} demonstrated remarkably low mosaicity, with peak widths as narrow as $\leq$ 60 $\mu$rad.

\subsection{The experimental setup}
\noindent T09 is an extracted beamline of the CERN Proton Synchrotron, capable of delivering a high purity ($> 90\%$) electron beam with an energy of up to 6 GeV.

The beamline was configured to deliver a 6 GeV electron beam with the smallest achievable angular divergence, a crucial parameter when dealing with directional effects that arises only if the particles impinge at a small angle with respect to the considered axis or planes. The measured angular divergence resulted to be about 1.5 mrad in the vertical plane and $\sim$ 1 mrad in the horizontal one. The main drawback of a low divergence beam is that the beam profile is broad, in our case it was around 10 cm. The intensity of electrons impinging onto the sample resulted to be of 10 - 100 particles per spill, with a spill being extracted every 30 - 60 s on average, even if the beamline itself would be capable of higher rates. Indeed, we set a lower rate to match the capability of our data acquisition system.

\begin{figure*}[h]
    \centering
    \includegraphics[width=0.8\linewidth]{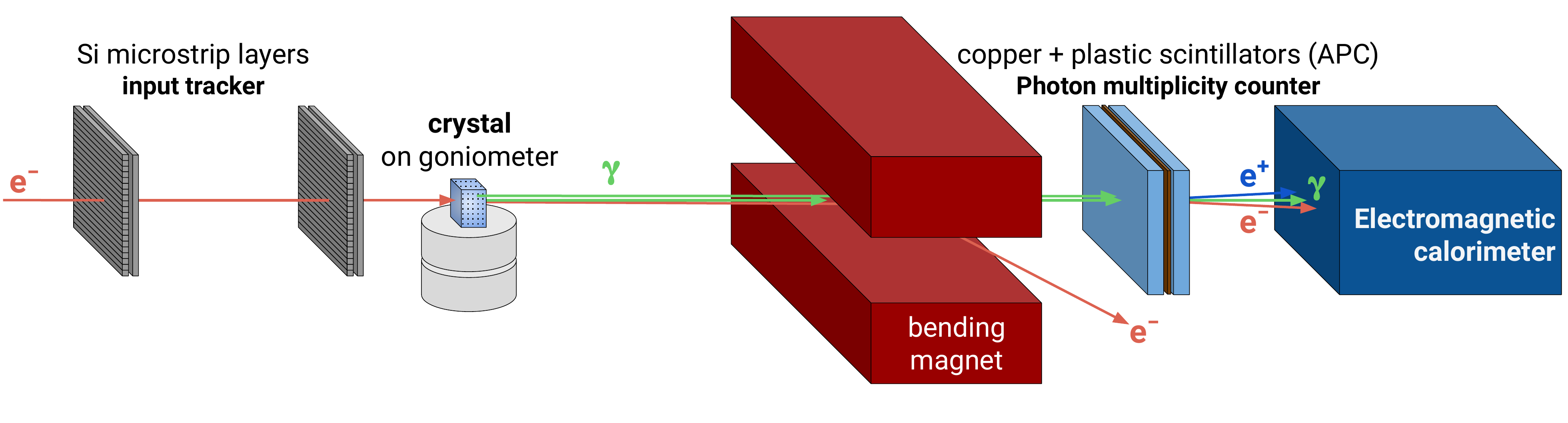}
    \caption{Sketch of the experimental setup used at CERN Proton Synchrotron. The electron beam (red) passes through two silicon trackers for particle trajectory reconstruction. The crystalline sample is mounted on a high-precision goniometer to ensure a precise alignment relative to the beam. A dipole magnet separates photons (green) from charged particles. The Active Photon Converter (APC) converts photons in a thin copper layer and detects the resulting $e^+e^-$ pairs (blue, red). A lead glass calorimeter measures the total energy loss by each primary electron.}
    \label{fig:3}
\end{figure*}

The experimental layout used to test the performance of a crystal as a photon radiator is presented in Fig. \ref{fig:3} and features the following key elements.

Two Čerenkov threshold detectors (not shown in figure \ref{fig:3}) were used to select only electrons among all the particles inside the beam.

To track the incoming particle, two silicon microstrip beam chambers with a sensitive area of 9.3 $\times$ 9.3 cm$^2$ were installed approximately 3 m apart from each other. These detector provided a spatial resolution of 40 $\mu$m \cite{AGILE} and an angular resolution of $\sim$ 200 $\mu$rad including the in-air multiple scattering contribution.

At about 0.7 m from the second (downstream) silicon tracker, the crystal sample was installed on a high-resolution multi-stage goniometer, to accurately set the crystal position and the relative angle between the beam incidence direction and the desired crystal axis. The goniometer stages can perform steps of 5 $\mu$m for the linear directions and of 1 $\mu$rad for the angular ones, which is well below the critical angle \linebreak of $\sim$ 540 $\mu$rad. 

To separate the produced photons from the charged particles, a 0.53 Tm dipole magnet was placed immediately downstream of the crystal.

\begin{figure}[h]
    \centering
    \includegraphics[width=0.75\linewidth]{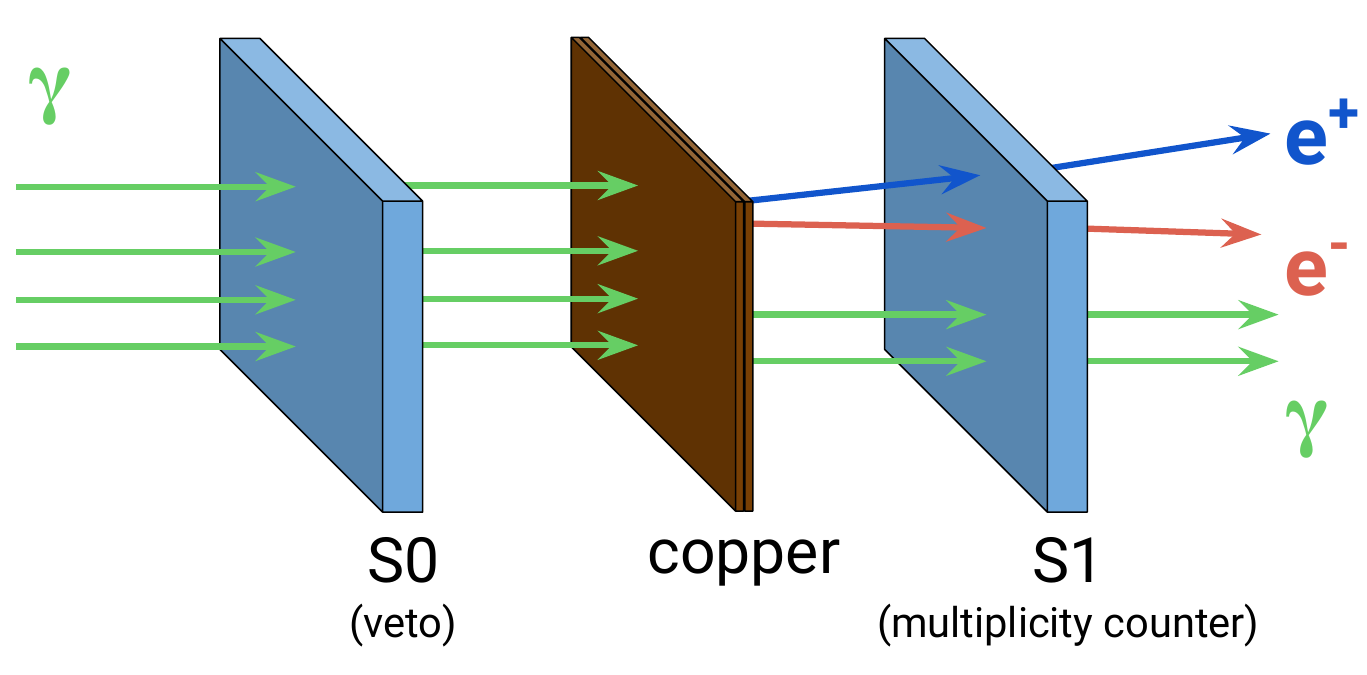}
    \caption{Active Photon Converter (APC) scheme. S0 scintillator act like a veto, while the S1 counts the multiplicity of the pair converted by the photons impinging on the copper thin layer.}
    \label{fig:4}
\end{figure}

Following, an Active Photon Converter (APC) was used, consisting of a thin copper layer (0.1 – 0.4 $X_0$) sandwiched between two 1 cm thick plastic scintillators. The first scintillator acts as a veto allowing the APC to discard the events in which not only photons are present. Within the copper layer, a small fraction of the photons may convert in an electron-positron pair which can be detected by the second plastic scintillator. This system, combined with simulation, provides an average information about the photon multiplicity, useful both to find the correct axial orientation and to measure the produced radiation \cite{EPJC22}. A schematic drawing of the APC is given in Fig. \ref{fig:4}

The experimental setup ended with a 37 cm long \linebreak($\sim$ 24.6 $X_0$) lead glass electromagnetic calorimeter with a transverse area of 10 $\times$ 10 cm$^2$ ~\cite{OPAL}, which measures the deposited energy by fully absorbing the produced radiation. The calorimeter was experimentally calibrated with beams of different energies in the range 1 - 6 GeV, thus determining the relationship between the analog-digital units of the digitizer and the nominal beam energy. Then, the calibration was refined through Monte Carlo simulations, using a dedicated simulation that reproduced the experimental setup. In this way the relationship between the nominal beam energy and the energy actually deposited in the calorimeter was determined.  

\section{Results}
\begin{figure}[h]
    \centering
    \includegraphics[width=1\linewidth]{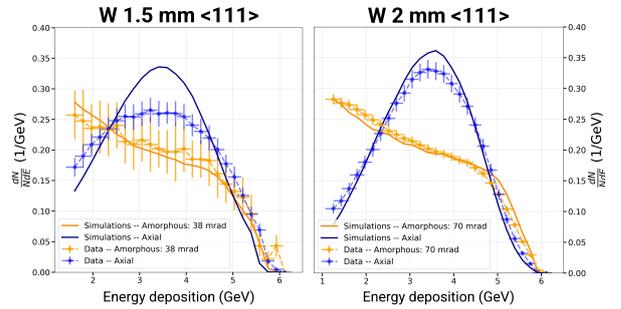}
   \caption{Radiated energy loss in the calorimeter for different crystals in axially oriented (blue) and amorphous oriented (orange) configurations. (a) W crystal, 1.5 mm thick, oriented along $\langle 111 \rangle$. (b) W crystal, 2 mm thick, oriented along $\langle 111 \rangle$. The experimental data are represented by dots and the simulation results by lines. The errors on the ordinates are statistical, while those on the abscissas are primarily driven by the energy resolution of the calorimeter.}
    \label{fig:5}
\end{figure}

\noindent The crystals were examined under various orientations relative to the incident beam. Fig. \ref{fig:5} presents the energy loss $E_\text{loss}$ spectra obtained for two distinct angular configurations for each of the two tested crystals: one aligned with the $\langle$111$\rangle$ crystal axis (blue dots) and the other at least 38 mrad away from the axis (orange dots). The latter configuration is far enough from the primary crystal axes for the atomic structure to effectively resemble a amorphous distribution. In this case, the orange curve exhibits the typical bremsstrahlung spectrum, characterized by a continuous distribution that gradually decreases as $E_\text{loss}$ increases. In contrast, the axial configuration (blue) reveals a significant suppression of the low-energy component and a pronounced enhancement at higher energies that reaches a maximum at approximately 3.5 GeV. 
The ratio between average axial and amorphous energy deposit in the different analysed samples, is summarized in Table \ref{tab:2}, where the results obtained with a research-centre tungsten crystal in \cite{EPJC22} are also reported for comparison.

In order to gain a deep insight of the obtained results, we performed full Monte Carlo simulations of the experiment. The simulations were performed using the Geant4 \cite{Geant4_0,Geant4_1,Geant4_2} toolkit embedding the new \texttt{G4ChannelingFastSimModel}\footnote{Additional information about the Channeling Fast Sim Model can be found at \href{https://geant4-userdoc.web.cern.ch/UsersGuides/PhysicsReferenceManual/html/solidstate/channeling/channeling_fastsim.html}{G4ChannelingFastSimModel}.} class \cite{Sytov19,Sytov23} that enable us to simulate the coherent interactions with the crystalline media.
The simulation of the radiation emission is based on the Monte Carlo integration of the general Baier-Katkov formula \cite{RADCHARM,Guidi_PRA_2012}. This approach leads to an accurate calculation of the radiation emission probability along the particle trajectories, taking automatically into account all the processes described in section \ref{sect:coherent_effects}.
To be able to perform this kind of simulations, the user has to pre-calculate quantities, such as the electric field and electronic/nuclear charge densities, that depend on the considered crystalline material, orientation and temperature. We did it for W $\langle111\rangle$ at room temperature and, in this case, the simulations closely match the experimental data provided that the crystal quality is sufficiently high to approximate the crystal as ideal. This is apparent in Fig. \ref{fig:5}b, which shows the results for a sample with a mosaicity about 1 mrad, as shown in Fig. \ref{fig:2}b. The agreement remains reasonable even for lower-quality crystals, as demonstrated in Fig. \ref{fig:5}a, where the mosaicity is of the order of several mrad, as depicted in Fig. \ref{fig:2}a. 

\begin{table}[h]
    \centering
    \resizebox{\columnwidth}{!}{
        \begin{tabular}{ccc}
        \hline
        Crystal & Distance                &  Axial/amorphous \\
        sample  & axis-amorphous (mrad) & average energy deposit\\
        \hline\hline
        W 2.25 mm $\langle100\rangle$ \cite{EPJC22} & 28 & $\sim$ 138\% \\
        \hdashline
        W 2 mm $\langle111\rangle$ & 70 & $\sim$ 135\% \\
        W 1.5 mm $\langle111\rangle$ & 38 & $\sim$ 119\% \\
        \hline
        \end{tabular}
    }
    \caption{Ratio between average axial and amorphous energy deposit in the different analysed samples. Results of \cite{EPJC22} are shown as reference. }
    \label{tab:2}
\end{table}

\begin{figure}[h]
    \centering
    \includegraphics[width=0.65\linewidth]{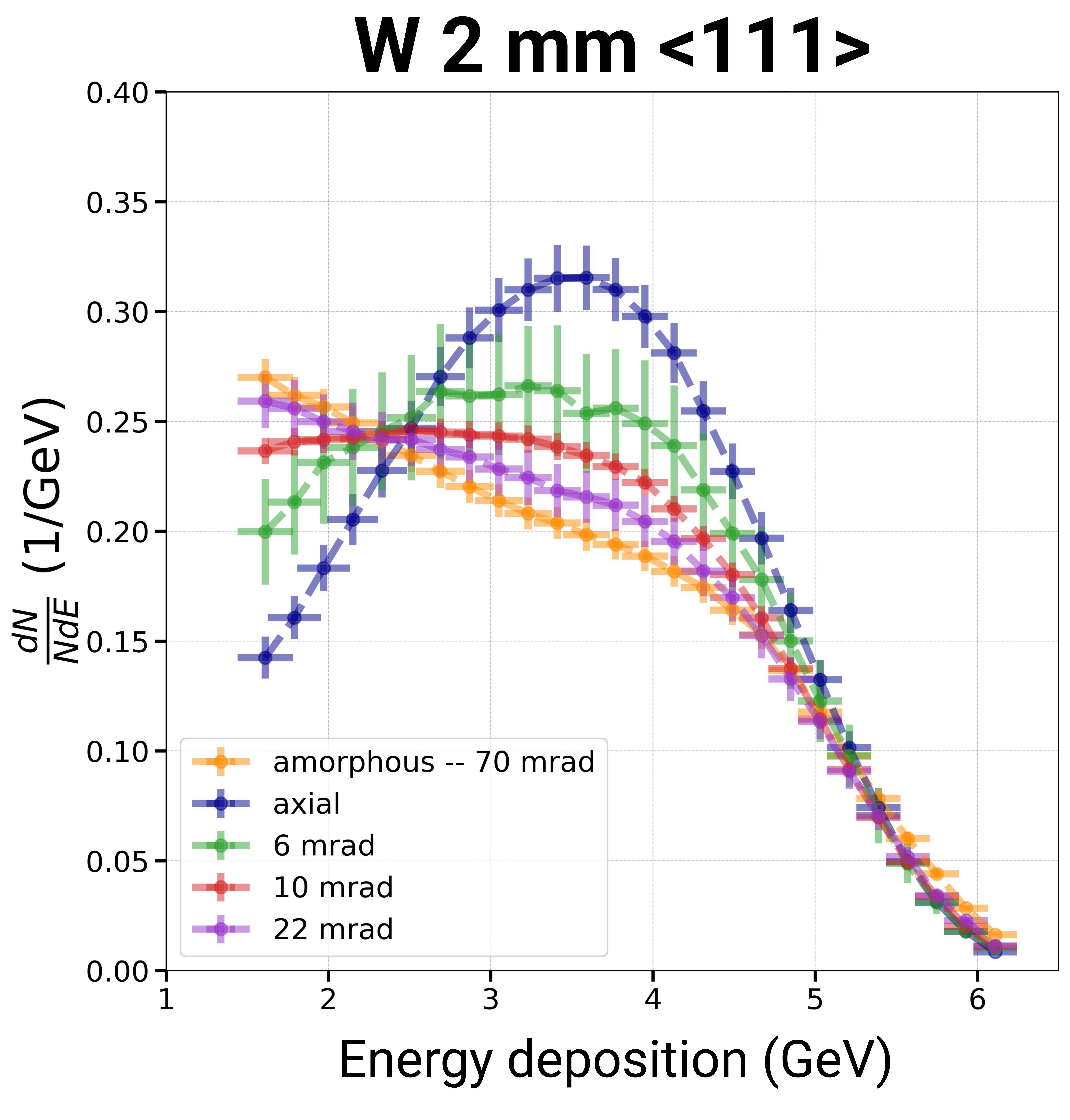}
    \caption{Radiated energy loss in the calorimeter for the transition scan showing the continuous transition from the amorphous to the aligned mode due to coherent effects for the 2 mm thick W sample. A noticeable enhancement is observed around $\sim$ 6-10 mrad (red dots). The errors on the ordinates are statistical, while those on the abscissas are primarily driven by the energy resolution of the calorimeter.}
    \label{fig:6}
\end{figure}

Fig. \ref{fig:6} presents a scan, moving from the aligned axial to the amorphous configuration. A continuous transition from the amorphous to the aligned mode with the axis can be observed, spanning an angular range of 15 mrad, which is significantly wider than the channeling critical angle ($\sim$ 0.54 mrad). This is due the aforementioned over-barrier effects and the contribution of coherent bremsstrahlung. Indeed, the latter does not involve a critical angle around the crystal axis to occur and contributes up to several mrad of misalignment.

\section{Conclusions and Outlook}
\noindent Industrial-grade radiators demonstrated robust performance despite their imperfections. The supply process can be then simplified, since the use of highly specialized research infrastructures is not strictly required. Additionally, thermal annealing has been shown to recover and improve crystal quality, enhancing their performance \cite{W_annealing}. The observed continuous transition from the amorphous to the aligned mode in tungsten radiators further highlights their potential for practical applications, as coherent effects remain significant even with industrial-grade crystals. This is particularly noteworthy given the wide use of amorphous targets as radiators.

These results open up new possibilities for various applications, including neutron production through photo-transmutation, radionuclide generation via photo-nuclear reactions and the development of novel positron sources for future lepton colliders, such as FCC-ee. 
The experimental results provided a benchmark for the simulation model, which was subsequently utilized in designing the FCC-ee positron source, culminating in an optimized single-crystal configuration \cite{ ALHARTHI25,Chaikovska_FCC2024}.

\section{Acknowledgement}
\noindent We acknowledge financial support under the National Recovery and Resilience Plan (NRRP), Call for tender No. 104 published on 02.02.2022 by the Italian Ministry of University and Research (MUR), funded by the European Union – NextGenerationEU – Project Title : "Intense positron source Based On Oriented crySTals - e+BOOST" 2022Y87K7X – CUP I53D23001510006.

This work was supported by the European Commission through the H2020-MSCA-RISE N-LIGHT (G.A. 872196), EIC-PATHFINDER-OPEN TECHNO-CLS (G.A. 101046458), MHz-TOMOSCOPY (G.A 101046448) and H2020-MSCA-IF-Global TRILLION (G.A.101032975) projects.

Iryna Chaikovska would like to thank the ANR (Agence Nationale de la Recherche) under Grant No. ANR-21-CE31-0007 for its support.

We also acknowledge partial support of INFN CSN1 (RD\_FCC and RD\_MUCOL projects) and CSN5 (OREO and Geant4INFN projects).

\bibliography{biblio.bib}

\end{document}